\documentclass[aip, apl, superscriptaddress, reprint, showpacs]{revtex4-1}

\usepackage{amsfonts}
\usepackage{amsmath}
\usepackage{amssymb, bm, mathrsfs}
\usepackage{graphicx, epstopdf, color}
\usepackage{soul}
\usepackage{cleveref, appendix}

\begin{document}
\title{Two-dimensional van der Waals electrical contact to monolayer MoSi$_2$N$_4$}

\author{Liemao Cao}
\affiliation{College of Physics and Electronic Engineering, Hengyang Normal University, Hengyang 421002, China}
\affiliation{Science, Mathematics and Technology (SMT), Singapore University of Technology and Design (SUTD), 8 Somapah Road, Singapore 487372, Singapore}

\author{Guanghui Zhou}
\affiliation{Department of Physics, Hunan Normal University, Changsha 410081, China}

\author{Qianqian Wang}
\affiliation{Science, Mathematics and Technology (SMT), Singapore University of Technology and Design (SUTD), 8 Somapah Road, Singapore 487372, Singapore}

\author{L. K. Ang}
\email[]{Authors to whom correspondence should be addressed: ricky$_$ang@sutd.edu.sg; and yeesin$_$ang@sutd.edu.sg}

\author{Yee Sin Ang}
\email[]{Authors to whom correspondence should be addressed: ricky$_$ang@sutd.edu.sg; and yeesin$_$ang@sutd.edu.sg}

\affiliation{Science, Mathematics and Technology (SMT), Singapore University of Technology and Design (SUTD), 8 Somapah Road, Singapore 487372, Singapore}

\begin{abstract}
Two-dimensional (2D) MoSi$_2$N$_4$ monolayer is an emerging class of air-stable 2D semiconductor possessing exceptional electrical and mechanical properties.
Despite intensive recent research efforts devoted to uncover the material properties of MoSi$_2$N$_4$, the physics of electrical contacts to MoSi$_2$N$_4$ remains largely unexplored thus far.
In this work, we study the van der Waals heterostructures composed of MoSi$_2$N$_4$ contacted by graphene and NbS$_2$ monolayers using first-principle density functional theory calculations.
We show that the MoSi$_2$N$_4$/NbS$_2$ contact exhibits an ultralow Schottky barrier height (SBH), which is beneficial for nanoelectronics applications.
For MoSi$_2$N$_4$/graphene contact, the SBH can be modulated via interlayer distance or via external electric fields, thus opening up an opportunity for reconfigurable and tunable nanoelectronic devices.
Our findings provide insights on the physics of 2D electrical contact to MoSi$_2$N$_4$, and shall offer a critical first step towards the design of high-performance electrical contacts to MoSi$_2$N$_4$-based 2D nanodevices.
\end{abstract}

\maketitle


The discovery of monolayer graphene \cite{Novoselov} has initiated tremendous experimental and theoretical efforts for the search of new two-dimensional (2D) materials with exotic physical properties and functionalities.
One important example is the family of 2D transition metal dichalcogenides (TMDCs), such as MoS$_2$ and WS$_2$, whose potential in nanoelectronic \cite{electron}, optoelectronic \cite{opto}, photonic \cite{photon}, valleytronic \cite{valley} and energy device \cite{energy} applications has revolutionized nanomaterial science and technology in recent years.
The potential of 2D TMDCs in industrial-grade 2D nanodevices is further boosted by the enormous design flexibility offered by vertical van der Waals heterostructures (VDWHs) in which physical properties can be custom-made by vertically stacking different 2D atomic layers \cite{VDW, Roche,Bafekry1,Bafekry2,Bafekry3}.

Beyond 2D TMDC, ultrathin transition metal nitride (TMN) have been actively explored in recent years \cite{TMN_rev}.
As TMN is a nonlayered material, the experimental synthesis of air-stable and large-area 2D TMN monolayer remains a formidable challenge \cite{xiao}.
In 2020, MoSi$_2$N$_4$, a TMN-based monolayer with no known 3D parent structure, has been successfully synthesized using chemical vapor deposition (CVD) method \cite{Hong}, which represents a milestone in the search of 2D TMN-based nanomaterial.
MoSi$_2$N$_4$, composed of a MoN$_2$ monolayer sandwiched by two Si-N bilayers, is an indirect bandgap semiconductor with excellent ambient air stability.
The intrinsic electron and hole mobilities are predicted to be 270 cm$^2$V$^{-1}$s$^{-1}$ and 1200 $^2$V$^{-1}$s$^{-1}$, respectively, which are substantially higher than that of MoS$_2$.
Density funcitonal theory (DFT) simulation has further revealed an expansive family of MA$_2$Z$_4$ monolayers (M $=$ early transition metal, e.g. Mo, W, and Nb; A $=$ Si or Ge, Z $=$ N, P or As), covering semiconducting, metallic and magnetic phases.
The discovery of MoSi$_2$N$_4$ and MA$_2$Z$_4$ monolayer family thus opens up an uncharted territory for the exploration of 2D-material-based device technology.

Although recent experiment have shed important lights on the structural, electrical, mechanical and optical properties of MoSi$_2$N$_4$ \cite{Hong}, the physics of electrically contacting MoSi$_2$N$_4$ with metals \cite{Venugopal,Yee,Wu,Banerjee,Xiong,Schulman,Allain} remains largely unknown thus far.
In 2D TMDC, their van der Waals (VDW) stacking with 2D metals has led to myriads of unusual characteristics, such as weak Fermi level pinning \cite{Liu_DFT}, tunable band gap \cite{Ebnonnasir, FZhang, Cao1}, miniband formation \cite{Pierucci}, and opto-valleytronic spin injection \cite{Luo}.
In relevance to MoSi$_2$N$_4$, the rich physical phenomena already observed in 2D-TMDC/2D-metal contacts immediately lead to the following open questions:
Can MoSi$_2$N$_4$ be integrated with 2D metals, such as graphene and NbS$_2$, to form structurally stable VDWHs?
What is the electronic properties of such heterostructures?
What types of contact, \emph{ohmic} or \emph{Schottky}, are formed in such electrical contacts?
Can the contact properties and types be engineered via external tuning knobs such as electric field and interlayer distances?

In this work, we address the above questions by performing a first-principle DFT simulation on the electronic properties of MoSi$_2$N$_4$ when it is vertically contacted by 2D metallic electrodes -- graphene and NbS$_2$ monolayers.
We find that MoSi$_2$N$_4$/NbS$_2$ Schottky contacts exhibits an ultralow Schottky barrier height (SBH) of $0.042$ eV, which is beneficial for room-temperature device operation.
For MoSi$_2$N$_4$/graphene, the SBH can be controlled by interlayer distance or by external electric field, thus allowing the contact to be reconfigured between $p$-type and $n$-type Schottky contact.
As the extended family of MA$_2$Z$_4$ contains myriads of 2D monolayers with exotic physical properties yet to be unearthed, our findings shall form a harbinger for designing efficient electrical contact and VDWHs based on MA$_2$Z$_4$ monolayer family.

\begin{figure}
\center
\includegraphics[scale=0.58]{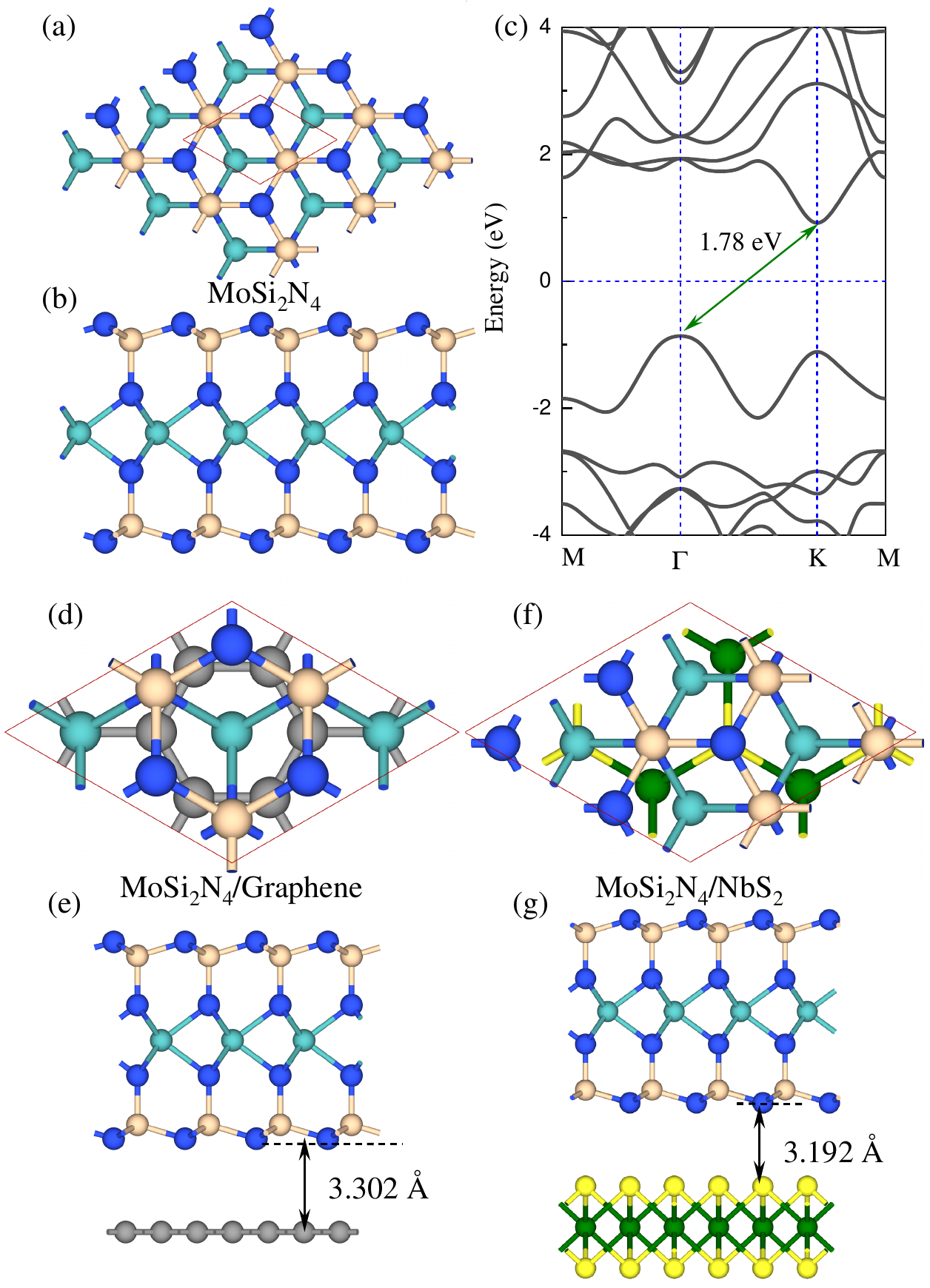}
\renewcommand{\figurename}{Fig. }
\caption{(Color online) (a) and (b) Top and side view of monolayer MoSi$_2$N$_4$. (c) Band structure of MoSi$_2$N$_4$. (d)-(g) Top and side view of optimized heterostructures of MoSi$_2$N$_4$/Graphene and MoSi$_2$N$_4$/NbS$_2$.}
\end{figure}

The structural optimization and electronic properties of an isolated monolayer MoSi$_2$N$_4$ and its VDWH when interfaced with graphene and NbS$_2$ are performed using the Vienna simulation package (VASP) \cite{Kresse,Kresse1}.
The projector-augmented wave (PAW) \cite{Kresse2} potentials are used to treat the electron-ion interaction.
The generalized gradient approximation (GGA) within Perdew-Burk-Ernzerhof (PBE) pseudopotentials are used to describe the exchange correlation functionals.
A cutoff energy of 500 eV is set.
DFT-D3 method is used to correct the effect of weak VDW interaction \cite{Grimme}.
A vacuum layer ($>20$ \AA) between the neighboring layers is employed along the z-direction of the heterostructures to eliminate the layer interactions caused by the neighboring slabs.
The break criterion for the electronic self-consistency is set to 10$^{-6}$ eV.
Monkhorst-Pack $k$-point grids of 9$\times$9$\times$1 are chosen.
All atoms are fully relaxed using the conjugated gradient method until the force is lower than 0.01 V/\AA.

The top and side views of an isolated MoSi$_2$N$_4$ are shown in Figs. 1(a) and (b), respectively.
The optimized lattice constant for MoSi$_2$N$_4$ is $a=b=2.91$ \AA, which is a hexagonal structure with a space group of $P6m1$.
MoSi$_2$N$_4$ is composed of septuple atomic layers with a thickness of 7.01 \AA.
Monolayer MoSi$_2$N$_4$ exhibits an indirect band gap of 1.78 eV [Fig. 1(c)], compared to that of MoS$_2$ ($1.9$ eV) and WS$_2$ ($2.0$ eV) monolayers \cite{TMDC_BG}. 
We further note that the calculations based on PBE functional is closer to the experimental values of 1.94 eV, when compared to the HSE06 values of 2.29 eV \cite{Hong} and 2.35 eV \cite{Bafekry}.

\begin{figure}
\center
\includegraphics[scale=0.3258]{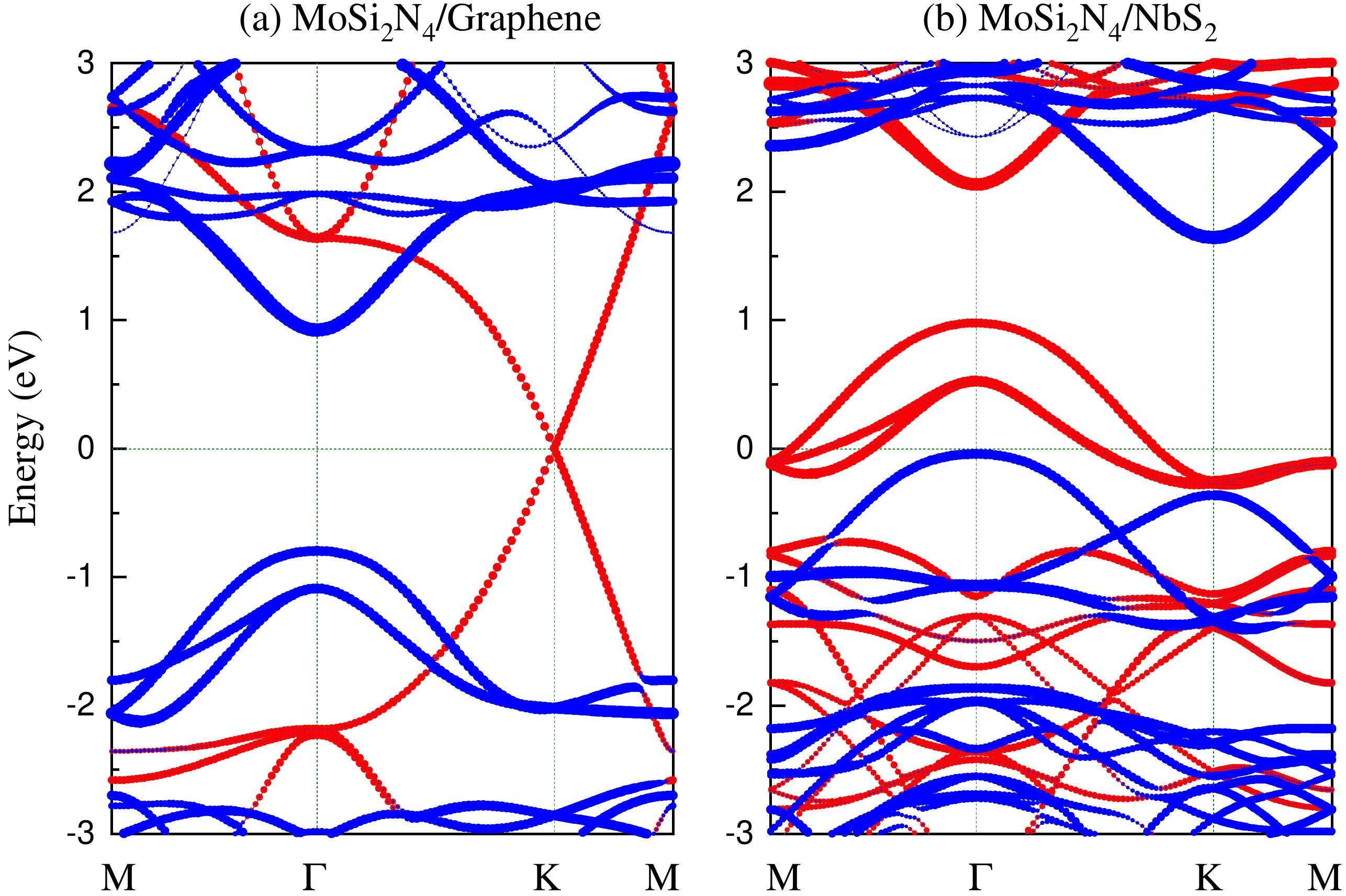}
\renewcommand{\figurename}{Fig.}
\caption{(Color online) The projected electronic band structures of (a) MoSi$_2$N$_4$/Graphene and (b) MoSi$_2$N$_4$/NbS$_2$. Here, blue and red symbols denote the contributions from theMoSi$_2$N$_4$ and graphene (NbS$_2$), respectively.}
\end{figure}

\begin{figure*}
	\center
	\includegraphics[scale=0.6058]{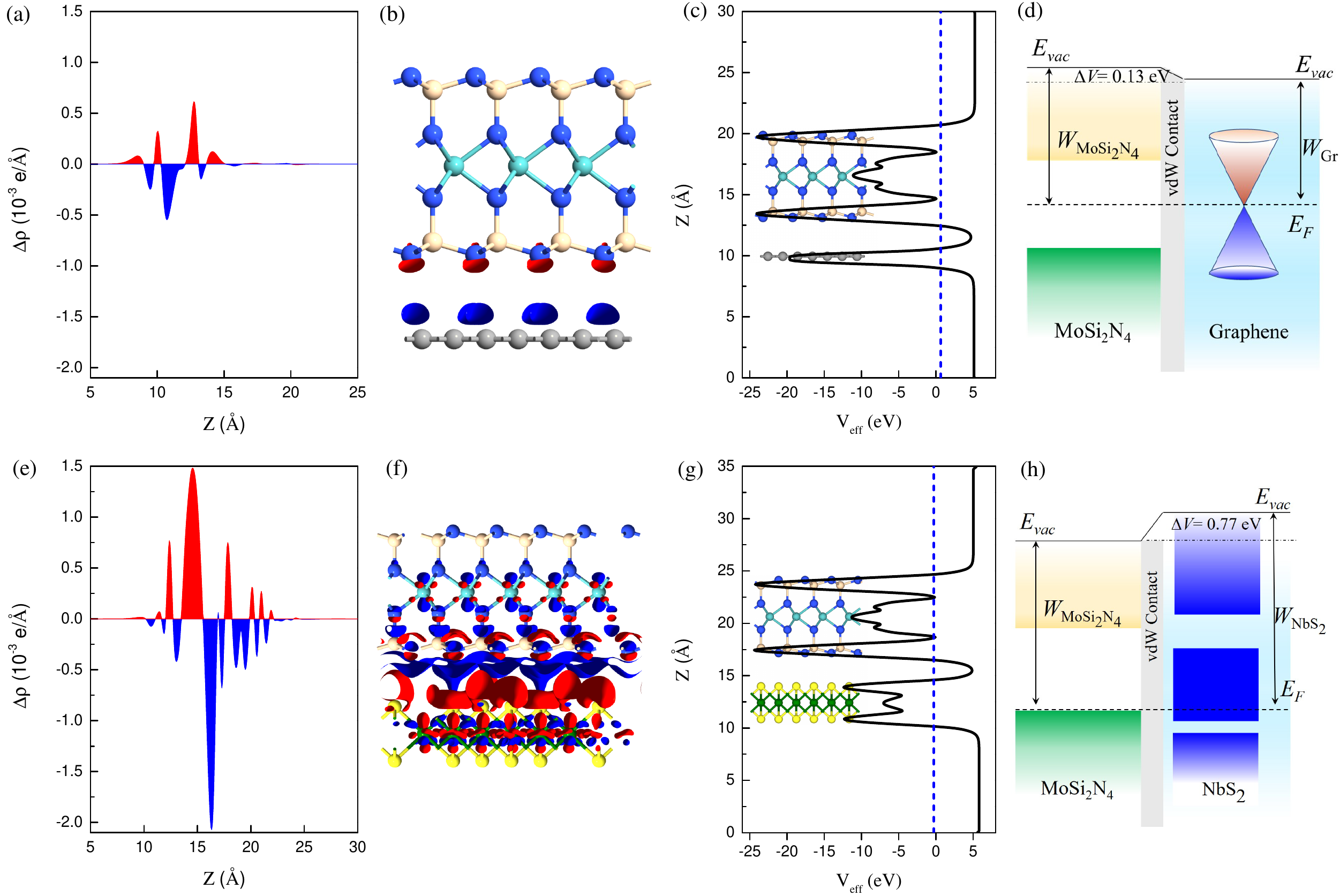}
	\renewcommand{\figurename}{Fig.}
	\caption{(Color online) Plane-averaged differential charge density$\Delta\rho$, 3D isosurface of the electron density difference, electrostatic potentials and band diagram of (a-d) MoSi$_2$N$_4$/Graphene and (e-h) MoSi$_2$N$_4$/NbS$_2$ heterostructures, respectively.}
\end{figure*}

We consider the 2D/2D contacts \cite{Cao1,Shin,Ding} formed by vertically contacting MoSi$_2$N$_4$ with graphene and NbS$_2$, namely MoSi$_2$N$_4$/graphene and MoSi$_2$N$_4$/NbS$_2$ VDWHs.
We use MoSi$_2$N$_4$ $\sqrt{3}\times\sqrt{3}$ and graphene 2$\times$2 supercells to form MoSi$_2$N$_4$/graphene.
For MoSi$_2$N$_4$/NbS$_2$, MoSi$_2$N$_4$ 2$\times$2 and NbS$_2$ $\sqrt{3}\times\sqrt{3}$ supercells are used [Fig. 1(d)-(g)].
To avoid the properties of MoSi$_2$N$_4$ being affected by mechanical stress, we fix the lattice of MoSi$_2$N$_4$ and apply strains in graphene and NbS$_2$ monolayer for MoSi$_2$N$_4$/graphene and MoSi$_2$N$_4$/NbS$_2$, respectively.
The lattice mismatch is 1.6\% and 0.07\% for MoSi$_2$N$_4$/graphene and MoSi$_2$N$_4$/NbS$_2$, respectively.
The stacking configurations have minimal effect on the physical properties of the heterostructures (see Supplementary Material).
After structural relaxation, the interlayer distances are obtained as 3.302 \AA$ $ and 3.192 \AA$ $ for MoSi$_2$N$_4$/graphene and MoSi$_2$N$_4$/NbS$_2$, respectively, which are much larger than the sum of the covalent radii of N and C (S) atoms.
The interfaces are found to be dominated by the VDW interactions.
We further test the interlayer distances using different VDW functionals (see Supplementary Materials), and the interlayer distances of the fully relaxed structures are found to be similar.
We calculated the binding energy $E_b$ as, $E_b = \left(E_H - E_M - E_{g(n)}\right)/N$ where $E_H$, $E_M$, $E_{g(n)}$, and $N$ denotes the total energies of the heterostructures, isolated MoSi$_2$N$_4$, isolated grephene (or NbS$_2$), and the number of atoms in the heterostructures, respectively.
The calculated binding energies are -199.89 eV/atom and -280.29 eV/atom for MoSi$_2$N$_4$/graphene, and MoSi$_2$N$_4$/NbS$_2$, respectively, which are substantially lower than other 2D-material-based heterostructures \cite{Padilha}.
Coupled with the excellent mechanical properties of MoSi$_2$N$_4$ \cite{Hong, Bafekry,C_yang}, the heterostructures are thus expected to be energetically favorable and stable.

The projected band structures of MoSi$_2$N$_4$/graphene and MoSi$_2$N$_4$/NbS$_2$ (Fig. 2) exhibit no obvious difference when SOC is included (see Supplementary Material).
The band structures of MoSi$_2$N$_4$, graphene and NbS$_2$ are well-preserved upon forming the heterostructures.
MoSi$_2$N$_4$ exhibits a direct band gap in MoSi$_2$N$_4$/graphene heterostructure due to band folding, which is a typical feature in 2D VDWHs.
We determine the SBH via the Schottky-Mott rule \cite{Bardeen}, i.e. $\Phi_n=E_{CBM}-E_F, \Phi_p=E_F-E_{VBM}$, where $\Phi_n$ and $\Phi_p$ are the interface potential barrier heights for electrons and holes, respectively; $E_{CBM}$, $E_F$ and $E_{VBM}$ are the energy of the conduction band minimum (CBM), Fermi energy, and valence band maximum (VBM).
For MoSi$_2$N$_4$/graphene, the $n$-type SBH ($\Phi_n)$ and the $p$-type SBH ($\Phi_p$) are 0.922 eV and $0.797$ eV, respectively, thus indicating the presence of a $p$-type Schottky contact.
In MoSi$_2$N$_4$/NbS$_2$, the $n$-type SBH ($\Phi_n$) and the p-type SBH ($\Phi_p$) is 1.642 eV and 0.042 eV, respectively.
The ultralow $p$-type SBH of MoSi$_2$N$_4$/NbS$_2$ contact immediately suggests the potential of NbS$_2$ as an efficient 2D electrical contact to MoSi$_2$N$_4$ with high charge injection efficiency particularly well-suited for room-temperature device applications.

We next examine the planar averaged charge density difference and electrostatic potentials of the MoSi$_2$N$_4$/graphene and MoSi$_2$N$_4$/NbS$_2$ heterostructures along the $z$ direction (Fig. 3).
We define the plane-averaged differential charge density ($\Delta\rho$) as $\Delta\rho=\rho_H-\rho_M-\rho_{g(n)}$, where $\rho_H$, $\rho_M$, and $\rho_{g(n)}$ are the plane-averaged charge density of the heterostructures, isolate plane-averaged charge densities of MoSi$_2$N$_4$ and graphene (NbS$_2$), respectively.
The symmetry of the MoSi$_2$N$_4$ lattice structure is broken when contacted by graphene and NbS$_2$, thus leading to charge redistribution.
Charge carriers are depleted around the graphene layer while accumulating near the MoSi$_2$N$_4$ layer [Figs. 3(a) and (b)].
In contrast, MoSi$_2$N$_4$/NbS$_2$ exhibits excess charges accumulate around the NbS$_2$ layer while depleting around the MoSi$_2$N$_4$ layer.
Moreover, the amount of transferred electrons between MoSi$_2$N$_4$ and NbS$_2$ is much greater than that of between MoSi$_2$N$_4$ and graphene by direct comparison between Figs. 3(a) and (e).
The 3D isosurfaces of the charge density difference is calculated in Figs. 3(b) and (f) for  MoSi$_2$N$_4$/graphene and MoSi$_2$N$_4$/NbS$_2$, respectively, which clearly illustrate the differences of charge transfer behavior in the two heterostructures.
Such charge transfer causes the Fermi level to move towards the valence band of MoSi$_2$N$_4$, resulting an ultralow SBH.
In Figs. 3(c) and (g), the plane-averaged electrostatic potential difference ($V_{eff}$) reveals that the direction of built-in electric field is different for the two heterostructures.
The corresponding band alignment diagrams of MoSi$_2$N$_4$/graphene and MoSi$_2$N$_4$/NbS$_2$ are shown in Figs. 3(d) and (h), respectively.

\begin{figure}
\center
\includegraphics[scale=0.598]{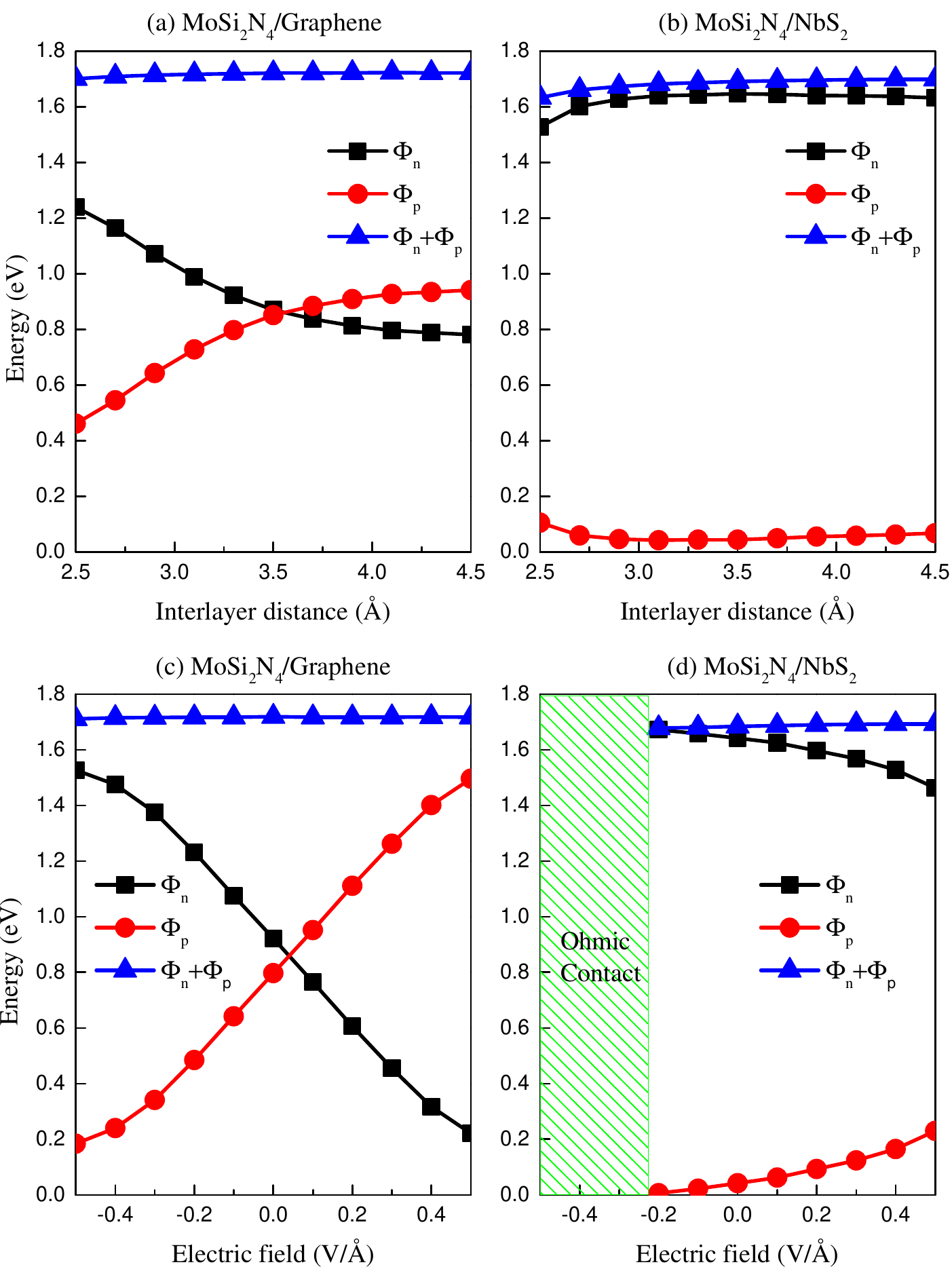}
\renewcommand{\figurename}{Fig.}
\caption{(Color online) SBH as a function of the interlayer spacing for (a) MoSi$_2$N$_4$/graphene and (b) MoSi$_2$N$_4$/NbS$_2$. SBH as a function of external electric field for (c) MoSi$_2$N$_4$/graphene and (d) MoSi$_2$N$_4$/NbS$_2$.}
\end{figure}

Changing the interlayer spacing, which can been achieved experimentally by nanomechanical pressure \cite{Dienwiebel}, diamond anvil cell \cite{Clark}, vacuum thermal annealing \cite{Tongay}, or by inserting hexagonal BN dielectric layers \cite{Fang}, provides a viable route to tune the electronic properties of a VDWHs.
The SBH of MoSi$_2$N$_4$/graphene and MoSi$_2$N$_4$/NbS$_2$ as a function of the vertical strain is illustrated in Fig. 4 [see Supplementary Material for their band structures].
Both electron and hole SBHs of MoSi$_2$N$_4$/graphene heterostructure are sensitively influenced by the interlayer distance [Fig. 4(a)]. For instance, an interlayer larger than 3.5 \AA$ $ signifies the transition from a $p$-type Schottky contact to an $n$-type Schottky contact.
In contrary, the SBH of MoSi$_2$N$_4$/NbS$_2$ heterostructure remains robust against interlayer distance variation [Fig. 4(b)].

Electrically tunable electronic properties is a useful characteristics for the design of all-electrical devices \cite{Padilha,Hu,Britnell}.
In Figs. 4(c) and (d), we examine the evolution of SBH as a function of an external electric field applied along the $z$-direction of the heterostructures (see Supplementary Material for the band structures).
We define the positive direction of the electric field as that pointing from MoSi$_2$N$_4$ layer to the 2D metals.
For MoSi$_2$N$_4$/graphene contact, when the external electric field is greater than 0.05 V/\AA, the contact is transformed from p-type to n-type Schottky contact [Fig. 4(c)].
The application of an external electric field thus provide a tuning knob to dynamically control the contact type of MoSi$_2$N$_4$/graphene heterostructure.
For MoSi$_2$N$_4$/NbS$_2$, the SBH is insensitive to an external electric field [Fig. 4(d)].
When a negative electric field exceeding -0.02 V/\AA$ $ is applied, the $p$-type SBH is completely eliminated, thus transforming MoSi$_2$N$_4$/NbS$_2$ into an ohmic contact.
The SBH tuning originates from the Fermi level shifting in graphene and NbS$_2$. When the interlayer spacing is changed or when an external electric field is applied, the balance of the system is destroyed. To re-establish equilibrium, the charges redistribute themselves across the interface which changes the Fermi level of contacting graphene and NbS$_2$ \cite{Bafekry4,Bafekry5} and, hence, their work functions. As a result, the SBH is modified.
\begin{figure}
\center
\includegraphics[bb=5 103 850 590, width=3.4 in]{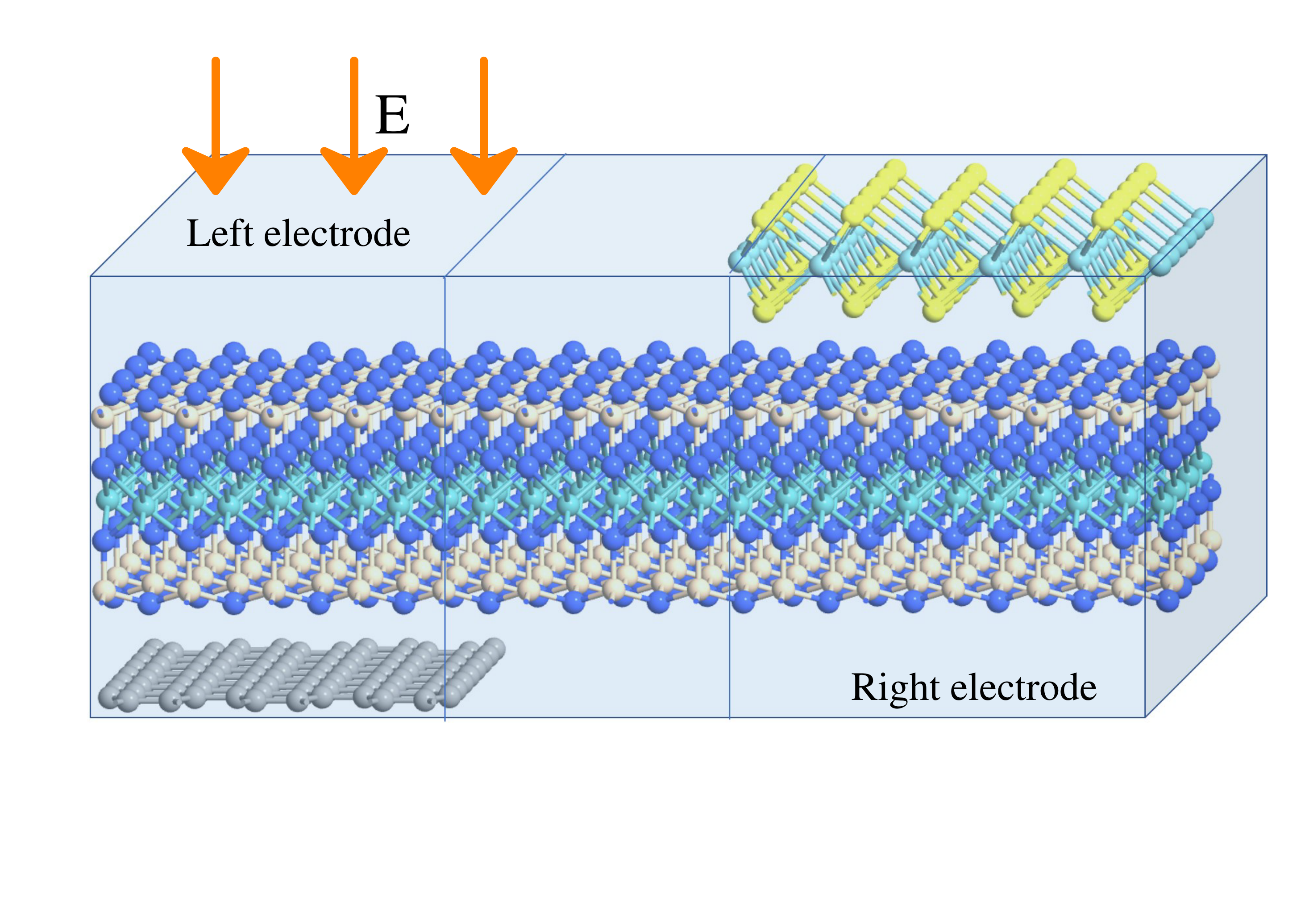}
\renewcommand{\figurename}{Fig.}
\caption{(Color online) Schematic diagram of a tunable nanodiode device based on a graphene/MoSi$_2$N$_4$/NbS$_2$ heterostructures.}
\end{figure}

Finally, we discuss a design of nanodiode composed of MoSi$_2$N$_4$/graphene and MoSi$_2$N$_4$/NbS$_2$ electrical contacts.
A 2D/2D \emph{homo-lateral} $p$-$n$ junction can be constructed using a single continuous sheet of MoSi$_2$N$_4$, contacted by graphene and NbS$_2$ on both sides (Fig. 5).
The MoSi$_2$N$_4$/NbS$_2$ region becomes $p$-doped, while the MoSi$_2$N$_4$/NbS$_2$ region serves as an electrically tunable contact.
By changing the electric field strength and polarity on the MoSi$_2$N$_4$/NbS$_2$, the device can be reconfigured between between $n$-$p$ or $n'$-$n$ nanodiodes.
The proposed nanodiodes is composed entirely of a single sheet of MoSi$_2$N$_4$ which is in contrast to the lateral\cite{li} or vertical \cite{Padilha1} heterostructure $p$-$n$ junction composed of two different 2D materials.
The lattice mismatch and interfacial carrier scattering effects are expected to be substantially reduced, thus improving the overall charge transport efficiency across the homojunction of the proposed device.

In summary, we investigated the electrical contacts between monolayer MoSi$_2$N$_4$ with 2D semimetal (graphene) and 2D metal (NbS$_2$), using firs-principle simulation.
Here, the ultralow SBH in MoSi$_2$N$_4$/NbS$_2$ contact can be a particularly beneficial feature for devices operating at room temperature.
For MoSi$_2$N$_4$/graphene contact, the SBH can be controlled externally via an electric field or via interlayer distance, thus revealing a potential in reconfigurable device applications.
Our findings shall shed important light on the design of MoSi$_2$N$_4$-based 2D/2D contacts and heterostructrues, thus paving a critical first-step toward the development of VDW heterostructure devices \cite{SJ} based on the emerging 2D material family of MA$_2$Z$_4$.

\section*{Supplementary Material}

See \textcolor{blue}{supplementary material} for the complete electronic band structures of MoSi$_2$N$_4$ heterostructures under different contact orientation, and the evolution of the electronic band structures under different interlayer distance and external electric fields. 

\begin{acknowledgments}
This work is supported by the Singapore MOE Tier 2 Grant (2018-T2-1-007).
LC acknowledge the supports of Hunan Provincial Natural Science Foundation of China (Grant No. 2019JJ50016), and science Foundation of Hengyang Normal University of China (No. 18D26).
All the calculations were carried out using the computational resources provided by the National Supercomputing Centre (NSCC) Singapore.

\textbf{Data Availability Statement.} The data that support the findings of this study are available from the corresponding author upon reasonable request.
\end{acknowledgments}

\end{document}